\documentstyle[12pt,psfig,preprint,aps]{revtex}
\makeatletter
\renewcommand{\@makefnmark}{\hbox{\mathsurround=0pt$^{\arabic{footnote})}$}}
\renewcommand{\@makefntext}[1]{\parindent=1em\noindent\hbox to 1.8em{\hss$^{\@thefnmark)}$}#1}
\def\@biblabel#1{#1.}
\makeatother

\newcommand{\ba}{\begin{eqnarray}}
\newcommand{\ea}{\end{eqnarray}}

\newcommand{\partialslash}{\partial\hspace{-.5em}/\hspace{.15em}}
\newcommand{\Pslash}{P\hspace{-.7em}/\hspace{.15em}}

\newcommand{\kslash}{k\hspace{-.5em}/\hspace{.15em}}

\newsavebox{\s}
\newsavebox{\tmpbox}
\newcommand{\Slash}[1]{\sbox{\s}{#1} \hbox to \wd\s {#1\hss\hbox to \wd\s{\hss/\hss}}}

\newcommand{\be}{\begin{equation}}
\newcommand{\ee}{\end{equation}}

\newcommand{\kstar}{K_{0}^{*}}
\newcommand{\diag}{{\rm diag}}

\newcommand{\Tr}{{\rm Tr}}
\newcommand{\mev}{{\kern2mm\rm MeV}}
\newcommand{\gev}{{\kern2mm\rm GeV}}

\newcommand{\journal}[5]{#1, {\it #2\/}, #4, {vol. #3}, p.~#5}
\newcommand{\journalX}[5]{#1, {\it #2\/}, #4}
\newcommand{\journalXP}[5]{#1, {\it #2\/}, #4, {vol. #3}, #5}
\newcommand{\journalXPV}[5]{#1, {\it #2\/}, #4,  #5}

\begin{document}

\begin{titlepage}
\title{Radial excitations of scalar and  $\eta$, $\eta'$ mesons
in a chiral quark model\\[1cm]
 \large\rm M.K.Volkov, V.L.Yudichev}
\end{titlepage}

\maketitle
\begin{abstract}
First radial excitations of the
scalar meson nonet and pseudoscalar mesons
$\eta$, $\eta'$ are described in a nonlocal chiral quark model of
the Nambu--Jona-Lasinio type with 't Hooft interaction. In this model
simple form factors are used, which allows us to describe first radial
excitations of the mesons and to conserve the gap equations describing
spontaneous breaking of chiral symmetry in the standard form.
The external parameters of form factors are fixed by  masses of
excited pseudoscalar mesons and the same form factors are used
for predicting the masses of excited scalar mesons.
Strong decays of excited scalar mesons and $\eta$, $\eta'$ mesons
are described in satisfactory agreement with experiment.
\end{abstract}
\vspace{1cm}

\sbox{\tmpbox}{Keywords:}
\newlength{\keyword}
\setlength{\keyword }{\textwidth}
\addtolength{\keyword}{-4cm}

\newpage

\section{Introduction}

In our previous papers \cite{weiss,volk97,ven,strange}
a nonlocal chiral quark model of the Nambu--Jona-Lasinio
(NJL) type was suggested to describe the first radial excitations of
mesons.
The nonlocality was introduced into the effective four-quark interaction
through form factors represented by first order polynomials of the
quark momentum squared $k^2$. In \cite{weiss} it was shown that such
form factors can be rewritten in  a relativistic form and the internal
parameters of these form factors (slope parameters)  can be chosen
so that the quark condensates and gap equations appearing
in the standard NJL model are unchanged.
With the form factors thus introduced,
all the low energy theorems are fulfilled in the chiral limit
(see \cite{weiss}).

In papers \cite{volk97,ven,strange}, this model was used for 
describing the mass spectrum of excited pions, kaons and of
the nonet of vector mesons. 
The main strong decays were also described therein.

Attempts to construct a model for describing radially
excited meson states were made, {\it e.~g.} in \cite{andr} where
a model with quasilocal four-quark interaction in the polycritical
regime was proposed. Different nonlocal models \cite{3p1,shakin,ned} were
also suggested.  

With the present work, we accomplish the invesigation of
the excited pseudoscalar meson nonet, considering the
excited states of $\eta$ and $\eta'$ and studying the first
radially excited states of the scalar meson nonet.

For a correct description of the $\eta$, $\eta'$
and  isoscalar scalar meson masses it is necessary,
in addition to the standard four-quark interaction,
to introduce six-quark 't Hooft interaction which breaks
the chiral symmetry and helps to solve the so called $U_A(1)$
problem. Contrary to the nonlocal four-quark interaction,
't Hooft vertices are pure local in accordance with
thier instantoneous origin.

The 't Hooft interaction gives rise to mixing of four
pseudoscalar states $\eta$, $\eta'$, $\hat\eta$ and $\hat\eta'$
(the caret symbol stands for radially excited meson states)
and four scalar states $\sigma$, $\hat\sigma$, $f_0$, $\hat f_0$.
After diagonalization of the free meson Lagrangians, we get the
mass spectrum of these meson states.

Each of the form factors we used has two arbitrary parameters:
the slope parameter $d_\alpha$ and external parameter $c_\alpha$.
There are three slope parameters --- $d_{uu}$, $d_{us}$, and $d_{ss}$.
They are uniquely defined by the condition that the excited
mesons do not  contribute to the quark condensate (tadpoles
including one form factor in the vertex equal to zero) and
therefore do not contribute to the gap equations.
Meanwhile, the constituent quark masses remain constant.
The external parameter $c_\alpha$ influences the interaction
of excited states of mesons with quarks
or  the corresponding four-quark interaction.
For the pseudoscalar and vector
mesons we define these parameters, using the masses of excited
states of mesons. However, for the scalar mesons we use
the same form factors as for the psudoscalar ones.
Thus, we can predict the masses of excited scalar meson states.
They turn out to  be in satisfactory agreement with experiment
and allow us to identify the members of scalar meson nonets and
tell us which of them are ground and which are  radially excited states.
This problem is discussed in the Conclusion.

After fitting the parameters $c_\alpha$ and  defining 
the basic model parameters  (constituent quark masses $m_u$ and
$m_s$ ($m_u\approx m_d$), ultraviolet
cut-off $\Lambda$, four-quark coupling constant $G$ and 't Hooft
coupling constant $K$), we can describe all the strong coupling constants
of mesons and calculate their strong decay widths.

Our paper is organized as follows. In Sec.~2, we introduce the
chiral quark Lagrangian with nonlocal four-quark vertices and
local 't Hooft interaction. In Sec.~3, we calculate the effective
Lagrangian for  isovector and strange mesons in the one-loop
approximation. There we renormalize meson fields and transform
the free part of the Lagrangian to the  diagonal form
and obtain meson mass formulae. Section 4 is devoted to  isoscalar mesons
where we find masses and mixing coefficients. The model parameters
are discussed in Sec.~5. In Sec.~6, we calculate the widths of
main strong decays of  excited states of
$a_0$,  $\sigma$, $f_0$ and $\kstar$ mesons.
In Sec.~7, we investigate strong decays of the
first radial excitations of $\eta$ and $\eta'$.
In Sec.~8, we analyze our results and compare them
with experimental data.
Some details of the calculations
fulfilled in Sec.~4 and 6 are given in Appendices A and B.

\section{$U(3)\times U(3)$ chiral Lagrangian with excited meson states
and 't Hooft interaction}
We use a  nonlocal separable four-quark interaction
of a current-current form which admits nonlocal vertices (form
factors) in the quark currents, and a pure local six-quark 't Hooft
interaction
\cite{klev,volk98}:
\ba
     {\cal L}(\bar q, q) &=&
     \int\! d^4x\; \bar q(x)
     (i \partialslash -m^0) q(x)+
     {\cal L}^{(4)}_{\rm int}+
     {\cal L}^{(6)}_{\rm int},  \label{lag}\\
     {\cal L}^{(4)}_{\rm int} &=&
     \frac{G}{2}\int\! d^4x\sum^{8}_{a=0}\sum^{N}_{i=1}
     [j_{S,i}^a(x) j_{S,i}^a(x)+
     j_{P,i}^a(x) j_{P,i}^a(x)],\\
     {\cal L}^{(6)}_{\rm int}&=&-K \left[\det
     \left[\bar q (1+\gamma_5)q\right]+
     \det\left[\bar q (1-\gamma_5)q\right]
     \right].
\ea
Here, $m^0$ is the current quark mass matrix ($m_u^0\approx m_d^0$) and
$j^a_{S(P),i}$ denotes the scalar (pseudoscalar) quark currents
\be
     j^a_{S(P),i}(x)=
     \int\! d^4x_1 d^4x_2\; \bar q(x_1)
     F^a_{S(P),i }(x;x_1,x_2) q(x_2)
\ee
where $ F^a_{S(P),i}(x;x_1,x_2)$ are the scalar (pseudoscalar)
nonlocal quark vertices.
To describe the first radial excitations of mesons, we take
the form factors in momentum space as follows (see \cite{weiss,volk97,ven}),
\be
\begin{array}{ll}
     F_{S,j}^a({\bf k})=\lambda^a
     f^a_j,\quad & F_{P,j}^a=
     i\gamma_5 \lambda^a f^a_j
\end{array}
\ee
\be
     f^a_1\equiv 1,\quad f^a_2\equiv f_a({\bf k})=c_a(1+d_a {\bf k}^2),\label{fDef}
\ee
where
$\lambda^a$ are Gell--Mann matrices,
$\lambda^0 = {\sqrt{2\over 3}}${\bf 1}, with {\bf 1} being the unit matrix.
Here, we consider the form factors in
the rest frame of mesons%
\footnote{The form factors depend on the transversal parts of
the relative momentum of quark-antiquark pairs $k_{\perp} =
k - \frac{k\cdot P}{P^2}P$, where $k$ and
$P$ are the relative and total momenta of a quark-antiquark pair,
respectively. Then, in the rest frame of mesons, ${\bf P}_{meson}$ = 0,
the transversal momentum is
$k_{\perp } = (0, {\bf k})$,
and we can define the form factors as depending on the 3-dimensional momentum ${\bf k}$ alone.
}%
.

The part of the Lagrangian (\ref{lag}), describing the ground states and
first radial excitations, can be rewritten in the following form
(see \cite{klev} and \cite{volk98}):
\ba
{\cal L}&=& \int\! d^4x \biggl\{
     \bar q(x) (i\partialslash-m^0) q(x) +
     \frac{G}{2}\sum_{a=0}^8 \left[\left(j_{S,2}^a\left(x\right)\right)^2+
     \left(j_{P,2}^a\left(x\right)\right)^2\right]+ \nonumber\\
&&   \frac12\sum_{a=1}^9\left[G^{(-)}_a
          \left(\bar q(x) \tau_a q(x)\right)^2+
     G^{(+)}_a\left(\bar q(x)i\gamma_5 \tau_a q(x)\right)^2\right]+
     \label{lagr}\\
&&   G_{us}^{(-)}(\bar q(x)\lambda_u q(x)) (\bar q(x)\lambda_s q(x)) +
     G_{us}^{(+)}(\bar q(x)i\gamma_5\lambda_u q(x))
          (\bar q(x)i\gamma_5\lambda_s q(x))\biggr\},\nonumber
\ea
where
\ba
     &&{\tau}_i={\lambda}_i ~~~ (i=1,...,7),~~~
     \tau_8 = \lambda_u = ({\sqrt 2}
     \lambda_0 + \lambda_8)/{\sqrt 3},\nonumber\\
&&   \tau_9 = \lambda_s = (-\lambda_0 +
     {\sqrt 2}\lambda_8)/{\sqrt 3}, \label{DefG} \\
&&   G_1^{(\pm)}=G_2^{(\pm)}=G_3^{(\pm)}=
     G \pm 4Km_sI_1(m_s), \nonumber \\
&&   G_4^{(\pm)}=G_5^{(\pm)}=G_6^{(\pm)}=
     G_7^{(\pm)}= G \pm 4Km_uI_1(m_u),
     \nonumber \\
&&   G_u^{(\pm)}= G \mp 4Km_sI_1(m_s), ~~~
     G_s^{(\pm)}= G, ~~~
     G_{us}^{(\pm)}= \pm 4{\sqrt 2}Km_uI_1(m_u).\nonumber
\ea
Here $m_u$ and $m_s$ are the constituent quark masses and $I_1(m_q)$
is the integral which for an arbitrary $n$ is defined as follows
\be
     I_n(m_q)={-i N_c\over (2\pi)^4}
     \int_{\Lambda_3}\!d^4 k
     {1\over (m^2_q-k^2)^n} .
     \label{DefI}
\ee
The 3-dimensional cut-off $\Lambda_3$ in (\ref{DefI})  is implemented
to regularize the divergent integrals%
\footnote{For instance,
$I_1(m)=\frac{N_c m^2}{8\pi^2}\left.[x\sqrt{x^2+1}-\ln(x+\sqrt{x^2+1})]\right|_{x=\Lambda_3/m}$.
}%
.

\section{The masses of isovector and strange mesons
(ground and excited states)}

After bosonization, the part of  Lagrangian (\ref{lagr}),
describing  the isovector and strange mesons, takes the form
\ba
&&   {\cal L}(a_{0,1},K_0^*{}_{,1},\pi_1,K_1, a_{0,2},  K_0^*{}_{,2}, \pi_2, K_2)=
     -\frac{a_{0,1}^2}{2G_{a_0}}-\frac{{K_0^*{}_{,1}}^2}{G_{K_0^*}}-\frac{\pi_1^2}{2G_\pi}-
     \frac{K_1^2}{G_K}-\nonumber\\
&&   \frac{1}{2G} ( a_{0,2}^2+2(K_0^*{}_{,2})^2+
     \pi_2^2+2 K_2^2)-\nonumber\\
    &&   i N_c \Tr\ln\left[1+
     \frac{1}{i\partialslash-m}\sum_{a=1}^7\sum_{j=1}^2 \lambda_a\left[
     \sigma^a_j+i\gamma_5\varphi^a_j\right]f^a_j
     \right]
\label{bosLag}
\ea
where $m=\diag(m_u,m_d,m_s)$ is the matrix of constituent quark masses
($m_u\approx m_d$),
$\sigma^a_j$ and $\phi^a_j$  are the scalar and pseudoscalar fields:
$\sum_{a=1}^{3}(\sigma^a_j)^2\equiv a_{0,j}^2=({a_{0,j}^0})^2+2a_{0,j}^+a_{0,j}^-$,
$\sum_{a=4}^{7}(\sigma^a_j)^2\equiv 2{\kstar{}_{,j}}^2=
2(\bar{\kstar}_{,j})^0(\kstar{}_{,j})^0+2(\kstar{}_{,j})^+(\kstar{}_{,j})^-$,
$\sum_{a=1}^{3}(\varphi^a_j)^2\equiv \pi_j^2=({\pi_j^0})^2+2\pi_j^+\pi_j^-$,
$\sum_{a=4}^{7}(\varphi^a_j)^2\equiv 2K_j^2=2\bar K^0_j K^0_j+2\bar K^+_jK^-_j$.
As to the coupling constants $G_a$, they
will be defined later (see Sect.~5 and (\ref{DefG})).

The free part of Lagrangian (\ref{bosLag}) has the following form
\ba
     {\cal L}^{(2)}(\sigma,\varphi)=\frac12\sum_{i,j=1}^2\sum_{a=1}^7
     \left(\sigma^a_i K_{\sigma,ij}^a(P)\sigma^a_j+
     \varphi^a_i K_{\varphi,ij}^a(P)\varphi^a_j\right)
     \label{L2}
\ea
    where the coefficients $K^{a}_{\sigma(\varphi), ij}(P)$ are given below,
\ba
    &&  K_{\sigma(\varphi), ij}^{a}(P)=
     -\delta_{ij}\left[\frac{\delta_{i1}}{G_a^{(\mp)}}+\frac{\delta_{i2}}{G}\right]-\nonumber\\
    &&   i N_c \Tr \int_{\Lambda_3}\frac{d^4k}{(2\pi)^4}
     {1\over \kslash+\Pslash/2-m^a_q}
     r^{\sigma(\varphi)}f_i^a
     {1\over \kslash-\Pslash/2-m^a_{q'}}
      r^{\sigma(\varphi)}f_j^a,
     \label{K_full}
\ea
\be
     r^\sigma=1,\quad r^{\phi}=i\gamma_5,
\ee
\be
     m_q^a = m_u~~(a = 1,...,7);\quad
     m_{q'}^a = m_u~~(a = 1,...,3);~~ m_{q'}^a = m_s~~ (a = 4,...,7),
     \label{m_q^a}
\ee
with $m_u$ and $m_s$ being the constituent quark masses and $f_j^a$ defined in
(\ref{fDef}).
Integral (\ref{K_full}) is evaluated by expanding in the
meson field momentum $P$. To order $P^2$, one obtains
\ba
     K_{\sigma(\varphi),11}^a(P)&=&
     Z_{\sigma(\varphi),1}^a (P^2 -
     (m_q^a\pm m_{q'}^a)^2- M_{\sigma^a(\varphi^a),1}^2 ),\nonumber\\
     K_{\sigma(\varphi),22}^a(P)
	 &=& Z_{\sigma(\varphi), 2}^a (P^2 -
     (m_q^a\pm m_{q'}^a)^2-  M_{\sigma^a(\varphi^a),2}^2 ),
     \nonumber \\
     K_{\sigma(\varphi),12}^a(P) &=& K_{\sigma(\varphi),21}^a(P) \;\; = \;\;
     \gamma_{\sigma(\varphi)}^a (P^2 - (m^a_q\pm m^a_{q'})^2 ),
     \label{Ks_matrix}
\ea
where
\be
     Z_{\sigma,1}^a = 4 I_2^a, \hspace{2em}  Z_{\sigma,2}^a  =  4 I_2^{ff a},
     \hspace{2em} \gamma_{\sigma}^a  = 4 I_2^{f a},
     \label{Zs}
\ee
\be
     Z_{\varphi,1}^a = Z Z_{\sigma,1}^a,
     \hspace{2em}  Z_{\varphi,2}^a  = Z_{\sigma,2}^a  ,
     \hspace{2em} \gamma_{\varphi}^a  = Z^{1/2}\gamma_{\sigma}^a
     \label{Zp}
\ee
and
\ba
      M_{\sigma^a(\varphi^a),1}^2 &=& (Z_{\sigma(\varphi),1}^a)^{-1}
     \left[\frac{1}{G_a^{(\mp)}}-4(I_1(m_q^a) +
     I_1(m_{q'}^a))\right]
     \label{M_1} \\
     M_{\sigma^a(\varphi^a),2}^2 &=& (Z_{\sigma(\varphi),2}^a)^{-1}
     \left[\frac{1}{G}-4(I_1^{ff a}(m_q^a) +
     I_1^{ff a}(m_{q'}^a))\right].
     \label{M_2}
\ea
The factor $Z$ here appears due to account of $\pi-a_1$-transitions~\cite{volk97,volk},
\be
Z=1-\frac{6 m_u^2}{ M_{a_1}^2},
\ee
and the integrals $I_2^{f..f}$  contain form factors:
\be
     I_2^{f..f_a}(m^a_q,m^a_{q'})={-i N_c\over (2\pi)^4}
     \int_{\Lambda_3} d^4 k
     {f_a({\bf k})..f_a({\bf k})\over ((m^a_q)^2-k^2)((m^a_{q'})^2-k^2)}.
     \label{DefIf}
\ee
Further, we consider only the scalar isovector and strange
mesons because the masses of the
pseudoscalar mesons have been already described in \cite{volk97}.

After the renormalization of the scalar fields
\be
     \sigma_i^{a r}=\sqrt{Z_{\sigma,i}^a} \sigma_i^{a}  \label{renorm}
\ee
the part of Lagrangian (\ref{L2}) which describes the scalar mesons
takes the form
\ba
     {\cal L}^{(2)}_{a_0}&=&\frac12
     \left(
     P^2-4 m^2_u -M^2_{a_0, 1}\right)a_{0, 1}^2+ \Gamma_{a_0}\left(
     P^2-4m_u^2\right)a_{0, 1}a_{0, 2}+\nonumber\\
     &&\frac12\left(P^2-4m_u^2- M_{a_{0},2}^2\right)a_{0, 2}^2,
     \label{La0}
\ea
\ba
     {\cal L}^{(2)}_{\kstar}\!&=&\!\frac12\! \left(
     P^2-(m_u+ m_s)^2\! -\! M^2_{\kstar,1}\right)\kstar{}_{,1}^2\!
     +\! \Gamma_{\kstar}
     \left(P^2-(m_u+m_s)^2\right)\kstar{}_{,1}\kstar{}_{,2}+\nonumber\\
     &&\!\frac12\left(P^2-(m_u+m_s)^2- M_{\kstar{},2}^2\right)\kstar{}_{,2}^2,
     \label{LK}
\ea
where
\be
     \Gamma_{\sigma^a}=\frac{I_2^{f_a}}{\sqrt{I_2 I_2^{ff_a}}}.\qquad
\ee
After the transformations of the meson fields
\ba
     \sigma^a
     &=& \cos( \theta_{\sigma,a} - \theta_{\sigma,a}^0) \sigma_1^{ar}
     - \cos( \theta_{\sigma,a} + \theta_{\sigma,a}^0) \sigma_2^{ar},   \nonumber \\
     \hat\sigma^{a}
     &=& \sin ( \theta_{\sigma,a} - \theta_{\sigma,a}^0) \sigma_1^{ar}
     - \sin ( \theta_{\sigma,a} + \theta_{\sigma,a}^0) \sigma_2^{ar},
     \label{transf}
\ea
Lagrangians (\ref{La0}) and (\ref{LK}) take the diagonal form:
\ba
     L_{a_0}^{(2)} &=& \frac12 (P^2 - M_{a_0}^2)~ a_0^2 +
     \frac12 (P^2 - M_{\hat a_0}^2)\hat a_0^{ 2}, \\
     L_{\kstar}^{(2)} &=& \frac12 (P^2 - M_{\kstar}^2)~ \kstar{}^2 +
     \frac12 (P^2 - M_{\hat\kstar}^2)\hat \kstar{}^{ 2}.
     \label{L_pK}
\ea
Here we have
\ba
&&   M^2_{(a_0, \hat a_0)} = \frac{1}{2 (1 - \Gamma^2_{a_0})}
     \biggl[M^2_{a_{0}, 1} + M^2_{a_{0}, 2}\pm \nonumber \\
&&   \qquad \sqrt{(M^2_{a_{0}, 1} - M^2_{a_{0}, 2})^2 +
     (2 M_{a_{0}, 1} M_{a_{0}, 2} \Gamma_{a_0})^2}\biggr]+4m_u^2, \\
&&   M^2_{(\kstar, \hat\kstar)} = \frac{1}{2 (1 - \Gamma^2_{\kstar})}
     \biggl[M^2_{\kstar, 1} + M^2_{\kstar, 2}\pm
       \nonumber \\
&&   \qquad  \sqrt{(M^2_{\kstar,1} - M^2_{\kstar,2})^2 +
     (2 M_{\kstar,1} M_{\kstar,2} \Gamma_{\kstar})^2}\biggr]+ (m_u+m_s)^2,
     \label{MpKstar}
\ea
and

\be
     \tan 2 {\bar{\theta}}_{\sigma,a} = \sqrt{\frac{1}{\Gamma_{\sigma^a}^2} -
     1}~\left[ \frac{M_{\sigma^a,1}^2-  M_{\sigma^a,2}^2}{M_{\sigma^a,1}^2
     + M_{\sigma^a,2}^2} \right],\qquad
     2 \theta_{\sigma,a} = 2 {\bar{\theta}}_{\sigma,a} + \pi,
     \label{tan}
\ee
\be
     \sin \theta_{\sigma,a}^{0} =\sqrt{{1+\Gamma_{\sigma^a}}\over 2}.
     \label{theta0}
\ee
The caret symbol stands for the first radial excitations of mesons.
Transformations (\ref{transf}) express the ``physical'' fields $\sigma$ and
$\hat\sigma$ through the ``bare'' ones $\sigma^{ar}_i$ and for calculations,
these equations must be inverted.
For practical use, we collect
 the values of the inverted equations for the scalar and
pseudoscalar fields%
\footnote{
Although the formulae for the pseudoscalars are not displayed here
(they  have been already obtained in \cite{volk97}) we need the
values because we are going to calculate the decay widths of processes
where pions and kaons are  secondary particles.
}
 in Table \ref{mixingTable}.


\section{The masses of isoscalar mesons (the ground and excited states)}

The 't Hooft interaction effectively gives rise to the additional
four-quark vertices in the isoscalar part of Lagrangian (\ref{lagr}):
\be
     {\cal L}_{\rm isosc}=\sum_{a,b=8}^9\left[
     (\bar q \tau_a q)T^S_{a b}
     (\bar q \tau_b q)+
     (\bar q i\gamma_5\tau_a q)T^P_{a b}
     (\bar q i\gamma_5\tau_b q)\right]
\ee
where $T^{S(P)}$ is a matrix with elements defined as follows
(for the definition of $G_u^{(\mp)}$, $G_s^{(\mp)}$ and $G_{us}^{(\mp)}$ see (\ref{DefG}))
\be
      \begin{array}{ll}
     T^{S(P)}_{88}=G^{(\mp)}_{u}/2,\quad & T^{S(P)}_{89}=G^{(\mp)}_{us}/2, \\
     T^{S(P)}_{98}=G^{(\mp)}_{us}/2,\quad & T^{S(P)}_{99}=G^{(\mp)}_{s}/2.
     \end{array}
\ee
This leads
to nondiagonal terms in the free part of the
effective Lagrangian for isoscalar
scalar and pseudoscalar mesons after bosonization
\ba
     &&{\cal L}_{\rm isosc}(\sigma,\varphi)=
	-{1\over 4}\sum_{a,b=8}^9\left[
     \sigma^{a}_1
     (T^S)^{-1}_{a b}\sigma^{b}_1 +
     \varphi^{a}_1
     (T^P)^{-1}_{a b}\varphi^{b}_1\right]-
\nonumber \\
     &&  {1\over 2G}\sum_{a=8}^9 \left[\left(\sigma^{a}_2\right)^2 +
     \left(\varphi^{a}_2\right)^2
     \right]-
\nonumber \\
     &&i~{\rm Tr}\ln \left\{1 + {1\over i{ \partialslash} - m}
	\sum_{a=8}^9\sum_{j=1}^2
	\tau^{a}[
     \sigma^{a}_j +
     i\gamma_5
     \varphi^{a}_j
     ]f^{a}_j \right\},  \label{Lbar}
\ea
where $(T^{S(P)})^{-1}$ is the inverse of $T^{S(P)}$:
\be
 \begin{array}{ll}
     (T^{S(P)})^{-1}_{88}=2G^{(\mp)}_{s}/D^{(\mp)},\quad &
	(T^{S(P)})^{-1}_{89}= (T^{S(P)})^{-1}_{98}=-2G^{(\mp)}_{us}/D^{(\mp)}, \\
	(T^{S(P)})^{-1}_{99}=2G^{(\mp)}_{u}/D^{(\mp)},\quad &
	D^{(\mp)}=G^{(\mp)}_u G^{(\mp)}_s-(G^{(\mp)}_{us})^2 .
 \end{array}
     \label{Tps1}
\ee
From (\ref{Lbar}), in the one-loop approximation, one obtains the
free part of the effective Lagrangian
\ba
     {\cal L}^{(2)}(\sigma,\phi)=\frac12\sum_{i,j=1}^2\sum_{a,b=8}^9
     \left(\sigma^a_i K_{\sigma,ij}^{[a,b]}(P)\sigma^b_j+
     \varphi^a_i K_{\varphi,ij}^{[a,b]}(P)\varphi^b_j\right).
	\label{Lisosc}
\ea
The definition of $K_{\sigma(\varphi),i}^{[a,b]}$ is given in
Appendix A.

After  the renormalization of both the scalar and pseudoscalar fields,
 analogous to (\ref{renorm}),  we come to the Lagrangian which
can be represented in a form slightly different from that
of (\ref{Lisosc}). It is convenient to
introduce 4-vectors of ``bare'' fields
\be
     \Sigma=(\sigma_{1}^{8\,r},\sigma_{2}^{8\,r},
	\sigma_{1}^{9\,r},\sigma_{2}^{9\,r}),\qquad
     \Phi=(\varphi_{1}^{8\,r},\varphi_{2}^{8\,r},
	\varphi_{1}^{9\,r},\varphi_{2}^{9\,r}).
\ee
Thus, we have
\ba
     {\cal L}^{(2)}(\Sigma,\Phi)=\frac12\sum_{i,j=1}^4
     \left(\Sigma_i {\cal K}_{\Sigma,ij}(P)\Sigma_j+
     \Phi_i {\cal K}_{\Phi,ij}(P)\Phi_j\right)
     \label{L2a}
\ea
where we introduced new functions ${\cal K}_{\Sigma(\Phi),ij}(P)$ (see Appendix A).

Up to this moment one has four pseudoscalar and four scalar meson states
which are the octet and nonet singlets. The mesons of the same parity
have the same quantum numbers and, therefore,
they are expected to be mixed. In our model the mixing is represented
by $4\times 4$ matrices $R^{\sigma(\varphi)}$ which
transform the ``bare'' fields 
$\sigma_{i}^{8\,r}$, $\sigma_{i}^{9\,r}$, $\varphi_{i}^{8\,r}$and $\varphi_{i}^{9\,r}$,
entering the 4-vectors $\Sigma$ and $\Phi$
to the ``physical'' ones 
 $\sigma$, $\hat\sigma$, $f_0$, $\hat f_0$,
$\eta$,  $\eta'$,  $\hat\eta$ and  $\hat\eta'$
 represented as components
of vectors
$\Sigma_{\rm ph}$ and $\Phi_{\rm ph}$:
\be
     \Sigma_{\rm ph}=(\sigma,\hat\sigma,f_0,\hat f_0),\qquad
     \Phi_{\rm ph}=(\eta,\hat\eta,\eta',\hat\eta')
\ee
where, let us remind once more,  a caret over a meson field
stands for the first radial excitation
of the meson. The transformation $R^{\sigma(\varphi)}$ is linear and nonorthogonal:
\be
     \Sigma_{\rm ph}=R^{\sigma}\Sigma,\qquad  \Phi_{\rm ph}=R^{\varphi}\Phi.
\ee
In terms of ``physical'' fields the free part of the effective
Lagrangian is of the conventional form and the coefficients
of matrices $R^{\sigma(\varphi)}$ give the mixing of
the $\bar uu$ and $\bar ss$ components, with and without form factors.

Because of the complexity of the procedure of diagonalization for the
matrices of dimensions greater than 2,
there is no such simple formulae as, {\it e.g.}, in (\ref{transf}).
Hence, we do not implement it analytically but
use numerical methods to obtain matrix elements
(see Table~\ref{isoscMixTab}).

\section{Model parameters and meson masses}

In our model we have five basic parameters: the masses of
the constituent $u(d)$ and $s$ quarks, $m_u=m_d$ and
$m_s$, the cut-off parameter $\Lambda_3$, the four-quark
coupling constant $G$ and the  't Hooft coupling constant
$K$. We have fixed these parameters with the help  of
input parameters: the pion decay constant $F_\pi=93 \mev$,
the $\rho$-meson decay constant $g_\rho=6.14$
(decay $\rho\to2\pi$)%
\footnote{Here we do not consider vector and axial-vector mesons,
however, we have used  the relation $g_\rho=\sqrt{6}g_{\sigma}$
 together with the Goldberger--Treiman relation
$g_\pi=\frac{m}{F_\pi}=Z^{-1/2}g_{\sigma}$ to fix the parameters
$m_u$ and $\Lambda_3$ (see \cite{volk97}).
},
the masses of pion and kaon and the mass difference of $\eta$ and $\eta'$
mesons (for details of these calculations, see \cite{volk97,ven,volk98}).
Here we give only numerical estimates of these parameters:
\ba
	&&m_u=280\mev,\quad m_s=405 \mev, \quad\Lambda_3=1.03 \gev, \nonumber\\
	&&G=3.14\gev^{-2},\quad K=6.1\gev^{-5}.
\ea
We also have a set of additional parameters $c^{\sigma^a(\varphi^a)}_{qq}$
in form factors $f^a_2$. These parameters are defined by masses of
excited pseudoscalar mesons,
 $c_{uu}^{\pi,a_0}=1.44$, $c_{uu}^{\eta,\eta',\sigma,f_0}=1.5$,
$c_{us}^{K,\kstar}=1.59$,
$c_{ss}^{\eta,\eta',\sigma,f_0}=1.66$.
The slope parameters $d_{qq}$ are fixed by special conditions satisfying
the standard gap equation,
$d_{uu}=-1.78 \gev^{-2}$,
$d_{us}=-1.76 \gev^{-2}$,  $d_{ss}=-1.73 \gev^{-2}$ (see \cite{volk97}).
Using these parmeters, we obtain masses of pseudoscalar and scalar mesons which
are listed in Table \ref{masses} together with experimental values.

From our calculations we come to the following interpretation of
$f_0(1370)$, $f_J(1710)$ and $a_0(1470)$ mesons: we consider them
as the first radial excitations of the ground states
$f_0(400-1200)$, $f_0(980)$ and $a_0(980)$. Meanwhile,
the meson $f_0(1500)$ is  likely a glueball. 
However, this is just our supposition. Only consideration
of a version of the NJL model with glueball states (or dilatons)
will allow us to clarify the status of $f_0(1500)$ and $f_0(1710)$.

\section{Strong decays of the scalar mesons}
The ground and excited states of scalar mesons $f_0$, $a_0$
decay mostly into  pairs of pseudoscalar
mesons. In the framework of a quark model and in the leading order
of $1/N_c$ expansion, the
processes are described by triangle quark
diagrams (see Fig.1).
Before we start to calculate the amplitudes, corresponding to
these diagrams,
we introduce, for convenience, Yukawa coupling constants
which naturally appear after
the renormalization (\ref{renorm}) of  meson fields:
\ba
	&&g_{\sigma_u}\equiv \left.g_{\sigma^a}\right|_{a=1,2,3,8}=[4I_2(m_u)]^{-1/2},
	\quad
	g_{\kstar}\equiv \left.g_{\sigma^a}\right|_{a=4,5,6,7}=[4I_2(m_u,m_s)]^{-1/2},
	\nonumber\\
	&&g_{\sigma_s}\equiv g_{\sigma^9}=[4I_2(m_s)]^{-1/2},
	\quad
	g_{\varphi^a}=Z^{-1/2}g_{\sigma^a}
	\nonumber\\
	&&g_{\pi}\equiv\left. g_{\varphi^a}\right|_{a=1,2,3},
	\quad
	g_{K}\equiv \left.g_{\varphi^a}\right|_{a=4,5,6,7},
	\quad
	g_{\varphi_u}\equiv g_{\varphi^8},
	\quad
	g_{\varphi_s}\equiv g_{\varphi^9}
     \label{g}
\ea
\ba
	&&\hat g_{\sigma_u}\equiv \left.\hat g_{\sigma^a}\right|_{a=1,2,3,8}=[4I_2^{ff}(m_u)]^{-1/2},
	\;\,
	\hat g_{\kstar}\equiv \left.\hat g_{\sigma^a}\right|_{a=4,5,6,7}=[4I_2^{ff}(m_u,m_s)]^{-1/2},
	\nonumber\\
	&&\hat g_{\sigma_s}\equiv \hat g_{\sigma^9}=[4I_2^{ff}(m_s)]^{-1/2},
	\quad
	\hat g_{\varphi^a}=\hat g_{\sigma^a}
	\nonumber\\
	&&\hat g_{\pi}\equiv\left.\hat g_{\varphi^a}\right|_{a=1,2,3},
	\;\,
	\hat g_{K}\equiv \left.\hat g_{\varphi^a}\right|_{a=4,5,6,7},
	\;\,
	\hat g_{\varphi_u}\equiv \hat g_{\varphi^8},
	\;\,
	\hat g_{\varphi_s}\equiv \hat g_{\varphi^9}
     \label{gX}
\ea

They can easily be related to $Z^a_{\sigma(\varphi),i}$ introduced in the
beginning of our paper.
Thus, the one-loop contribution to the effective Lagrangian can be
rewritten in terms of the renormalized fields:
\ba
  {\cal L}_{\rm 1-loop}(\sigma,\varphi)&=&
      i N_c \Tr\ln\left[1+
     \frac{1}{i\partialslash-m}\sum_{a=1}^9 \tau_a\left[
     g_{\sigma^a}\sigma^a_1+i\gamma_5 g_{\varphi^a}\varphi^a_1+\right.\right.\nonumber\\
	&&\left.(\hat g_{\sigma^a}\sigma^a_2+i\gamma_5 \hat g_{\varphi^a}\varphi^a_2)f_a\right]
     \Biggr]
\label{bosLag1}
\ea

All amplitudes that describe processes of the type $\sigma\to\varphi_1\varphi_2$
can be divided into two parts:
\ba
	T_{\sigma\to\varphi_1\varphi_2}&=&
	C\left(-\frac{i N_c}{(2\pi)^4}\right)
	\int_{\Lambda_3}d^4 k \frac{\Tr[(m+\Slash{k}+\Slash{p}_1)\gamma_5
	(m+\Slash{k})\gamma_5(m+\Slash{k}-\Slash{p}_2)]}{
	(m^2-k^2)(m^2-(k+p_1)^2)(m^2-(k-p_2)^2)}\nonumber\\
	&=& 4mC\left(-\frac{i N_c}{(2\pi)^4}\right)
	\int_{\Lambda_3}d^4k
	\frac{\left[1-\displaystyle\frac{p_1\cdot p_2}{m^2-k^2}\right]}{(m^2-(k+p_1)^2)(m^2-(k-p_2)^2)}
	\nonumber\\
	&=&4 m C [I_2(m,p_1,p_2)-p_1\cdot p_2 I_3(m,p_1,p_2)]=T^{(1)}+T^{(2)}
	\label{T}
\ea
here $C=4 g_{\sigma} g_{\varphi_1}g_{\varphi_2}$ and $p_1$, $p_2$ are
momenta of the pseudoscalar mesons.
Using (\ref{g}) and (\ref{gX}), we rewrite the amplitude
$T_{\sigma\to\varphi_1\varphi_2}$
in another form
\ba
&&T_{\sigma\to\varphi_1\varphi_2}\approx
4mZ^{-1/2}g_{\varphi_1} \left[1-p_1\cdot p_2
\frac{I_3(m)}{I_2(m)}\right],\label{T1}\\
&&p_1\cdot p_2 =\frac12(M_\sigma^2-M_{\varphi_1}^2-M_{\varphi_2}^2).
\ea
We assumed here that the ratio of $I_3$ to $I_2$  slowly changes with momentum
in comparison with factor $p_1\cdot p_2$,
therefore, we ignore their momentum dependence in (\ref{T1}).
With this assumption we
are going to obtain just a qualitative picture for  decays of the
excited scalar mesons.

In eqs. (\ref{T}) and (\ref{T1}) we omitted the contributions
from the diagrams which include form factors in vertices.
The whole set of diagrams consists of those containing zero, one, two
and three form factors. To obtain the complete amplitude, one must
sum up all contributions.

After these general comments, let us  consider the decays of
$ a_0(1450)$, $f_0(1370)$ and $f_J(1710)$.
First, we estimate the
decay width of the process $\hat a_0\to\eta\pi$,
taking the mixing coefficients from Table~\ref{mixingTable} and~\ref{isoscMixTab}
(see Appendix B for the details).
The result is
\be
      T^{(1)}_{\hat a_0\to\eta\pi}\approx0.2\gev,
\ee
\be
     T^{(2)}_{\hat a_0\to\eta\pi}\approx3.5 \gev,
\ee
\be
     \Gamma_{\hat a_0\to\eta\pi}\approx
     160 \mev.
\ee

From this calculation one can see that $T^{(1)}\ll T^{(2)}$ and the
amplitude is dominated by its second part, $T^{(2)}$, which is
momentum dependent. The first part is small because the diagrams
with different numbers of form factors cancel each other. As a consequence,
in all processes where an excited scalar meson decays into a pair of
ground pseudoscalar states, the second part of the amplitude
defines the rate of the process.

For the decay $\hat a_0\to\pi\eta'$ we obtain the amplitudes
\be
     T^{(1)}_{\hat a_0\to\pi\eta'}\approx0.8 \gev,
\ee
\be
     T^{(2)}_{\hat a_0\to\pi\eta'}\approx3 \gev,
\ee
and the decay width
\be
    \Gamma_{\hat a_0\to\pi\eta'}\approx36 \mev.
\ee
The decay of $\hat a_0$ into kaons is described by the amplitudes $T_{\hat a_0\to K^+K^-}$
and  $T_{\hat a_0\to \bar K^0K^0}$
which, in accordance with our scheme,  can again be  divided into two parts: $T^{(1)}$
and $T^{(2)}$ (see Appendix B for details):
\be
T_{\hat a_0\to K^+K^-}^{(1)}\approx 0.2\gev,
\ee
\be
T_{\hat a_0\to K^+K^-}^{(2)}\approx 2.1\gev.
\ee
and the  decay width  is
\be
\Gamma_{\hat a_0\to KK}=\Gamma_{\hat a_0\to K^+K^-}+\Gamma_{\hat a_0\to \bar K^0K^0}\approx 100\mev.
\ee
Qualitatively, our results do not contradict the experimental data.
\be
	\Gamma^{\rm tot}_{\hat a_0}=265\pm13 \mev,\quad BR(\hat a_0\to KK):BR(\hat a_0\to\pi\eta)= 0.88\pm0.23.
\ee
The decay widths
of radial excitations of scalar isoscalar mesons
are estimated in the same way as it was shown above. We obtain:
\be
\Gamma_{\hat\sigma\to\pi\pi}=\left\{
\begin{array}{l}
550 \mev (M_{\hat\sigma}=1.3 \gev) \\
460 \mev (M_{\hat\sigma}=1.25 \gev),
\end{array}
\right.
\ee
\be
\Gamma_{\hat\sigma\to\eta\eta}=\left\{
\begin{array}{l}
24 \mev (M_{\sigma}=1.3 \gev) \\
15 \mev (M_{\sigma}=1.25 \gev),
\end{array}
\right.
\ee
\be
\Gamma_{\hat\sigma\to\sigma\sigma}=\left\{
\begin{array}{l}
6 \mev (M_{\sigma}=1.3 \gev) \\
5 \mev (M_{\sigma}=1.25 \gev),
\end{array}
\right.
\ee
\be
\Gamma_{\hat\sigma\to KK}\sim 5 \mev,
\ee
\be
\begin{array}{lclclcl}
\Gamma_{f_0(1710)\to 2\pi}& \approx& 3\mev, &\quad &
     \Gamma_{f_0(1500)\to 2\pi}& \approx& 3\mev,\\
\Gamma_{f_0(1710)\to 2\eta}&  \approx& 40\mev, &\quad &
     \Gamma_{f_0(1500)\to 2\eta}& \approx& 20\mev,\\
\Gamma_{f_0(1710)\to \eta\eta'}&  \approx& 42\mev, &\quad &
     \Gamma_{f_0(1500)\to \eta\eta'}& \approx& 10\mev,\\
\Gamma_{f_0(1710)\to KK}&  \approx& 24\mev, &\quad &
     \Gamma_{f_0(1500)\to KK}& \approx& 20\mev.
\end{array}
\ee
The decays of $f_0(1500)$ and $f_0(1710)$ to $\sigma\sigma$ are negligibly small,
so we disregard them.

Here we desplayed our estimates  for both $f_J(1710)$ and $f_0(1500)$
resonances. Comparing them will allow us to decide which one to
consider as the first radial excitation of $f_0(980)$ and which a glueball.
From the experimental data:
\be
\Gamma^{\rm tot}_{\sigma'}=200 - 500 \mev,
\quad
\Gamma^{\rm tot}_{f_0(1710)}=133\pm 14 \mev,
\quad
\Gamma^{\rm tot}_{f_0(1500)}=112\pm 10 \mev
\ee
we can see that in the case of $f_0(1500)$ being a $\bar qq$ state
there is deficit in the decay widths whereas for $f_J(1710)$ the result is
close to experiment.
From this we conclude that  the meson
$f_J(1710)$ is a radially excited partner for $f_0(980)$
and the meson state $f_0(1370)$ is the
first radial excitation  of $f_0(400-1200)$.
As to the state $f_0(1500)$, we are inclined to consider it as
a glueball which
significantly contributes to the decay width.%
\footnote{
Let us emphasize again that it is only our preliminary conclusion. 
A more careful investigation of this problem will be
done in our further works.
}.

The first radially excited state of the strange scalar $\hat\kstar$
decays mostly to $K\pi$ and is characerazed by the width
\be
\Gamma_{\hat\kstar\to K\pi}\approx 300 \mev.
\ee
This value is in agreement with experiment:
\be
\Gamma^{exp}_{\kstar(1430)\to K\pi}\approx 287\pm23 \mev.
\ee

The strong decays widths of the ground states of scalar mesons
were calculated in paper \cite{volk98} in the framework of 
the standard NJL model with 't Hooft interaction where it was
shown that a strange scalar meson state with mass about $960\mev$ 
decays into $K\pi$ with the rate
\ba
&&\Gamma_{\kstar(960)\to K\pi}=
\frac{3}{Z\pi M_{\kstar}}\left(\frac{m_u m_s}{2F_\pi}\right)^2\times\nonumber\\
&&\qquad\sqrt{\frac{(M_{\kstar}^2-(M_K-M_\pi)^2)(M_{\kstar}^2-(M_K+M_\pi)^2)}{M_{\kstar}^4}}
\approx360 \mev
\ea
From comparing this result with the analisys of phase shifts given in
\cite{ishida} where  an evidence for existence of a scalar strange
meson with the mass equal to $905\pm50\mev$ and decay width 
$545\pm170\mev$ is shown, we identify the state $\kstar(960)$ as
a member of the ground scalar meson nonet. The state $\kstar(1430)$
is thereby its first radial excitation.

\section{Strong decays of $\eta(1295)$ and $\eta(1440)$.}

The mesons $\eta(1295)$ and $\eta(1440)$ have common
decay modes: $a_0\pi$, $\eta\pi\pi$, $\eta(\pi\pi)_{\sl S-wave}$,
$K\bar K\pi$,
moreover, the heavier pseudoscalar $\eta(1440)$ decays also into
$KK^*$. For the processes with two secondary particles,
the calculations of decay widths are done in the same way as
shown in the previous section, by calculating triangle diagrams
similar to that in Fig.~1.

Let us consider the decay $\hat\eta\to a_0\pi$. The corresponding
amplitude is of the same form  as given in (\ref{T}) for
decays of the type $\sigma\to\varphi_1\varphi_2$.
It can also be divided into two parts $T^{(1)}$ and  $T^{(2)}$
which in our approximation
are constant and momentum-dependent in the sense explained in
the previous section (see (\ref{T1}) and the text below):
\be
T^{(1)}_{\hat\eta\to a_0\pi}\approx 0.3 \gev,
\ee
\be
T^{(2)}_{\hat\eta\to a_0\pi}\approx -1  \gev.
\ee
Therefore, the decay width is
\be
\Gamma_{\hat\eta\to a_0\pi}\approx 3 \mev.
\ee

The decay $\hat\eta\to \eta(\pi\pi)_{\sl S-wave}$ is nothing else than
the decay $\hat\eta\to\eta\sigma\to\eta(\pi\pi)_{\sl S-wave}$ where
we have the $\sigma$-meson in the final state decaying then into
pions in the S-wave. We simply calculate $\hat\eta\to\eta\sigma$,
with $\sigma$ as a decay product.

The calculation of decay widths for the rest of the decay modes with
two particles in the final state is similar and the result
is given in Table~\ref{etadecay}.

The decay $\hat\eta'\to KK^*$ differs from the other modes
by a strange vector meson among the decay products.
In this case we have
\ba
     T_{\hat\eta'\to KK^*}^\mu&=&
     4(p_1+p_2)^\mu \biggl([g_{u}g_{K}g_{K^*}I_2(u,s)+\ldots]-\\
     &&\sqrt{2}[g_{s}g_{K}g_{K^*}I_2(u,s)+\ldots]\biggr)
\ea
where $p_1$ is the momentum of $\hat\eta'$; $p_2$, the momentum of
$K$; and dots stand for the terms with form factors (not displayed
here). 
These two parts are of the same order of magnitude and differ in 
sign and therefore cancel each other, which 
reduces the decay width up to tens of keV:
\be
\Gamma_{\hat\eta'\to KK^*}\approx 70~{\rm keV}.
\ee

When there are three particles in the final state, poles appear
in amplitudes, related to intermediate scalar
resonances. As it is well known from $\pi\pi$  scattering,
these diagrams can play a crucial role in the description of
such processes.
So, in addition to the "box" diagram we take into account the  diagrams with
poles provided by $\sigma$, $f_0$, and $a_0$ resonances
(see Fig.~2 for the decay $\hat\eta\to\eta\pi\pi$).
Here we neglect the momentum dependence 
in the box diagram, approximating it by a constant.
The amplitude is thereby
\ba
&&T_{\hat\eta\to\eta\pi\pi}=
B+{c_{\sigma\eta\hat\eta} c_{\sigma\pi\pi} \over
     M_{\sigma}-s-i M_{\sigma}\Gamma_{\sigma}   }+
{c_{f_0\eta\hat\eta} c_{f_0\pi\pi} \over
     M_{f_0}-s-i M_{f_0}\Gamma_{f_0}   }+\nonumber\\
&&{c_{a_0\hat\eta\pi} c_{a_0\eta\pi} \over
     M_{a_0}-t-i M_{a_0}\Gamma_{a_0}   }+
{c_{a_0\hat\eta\pi} c_{a_0\eta\pi} \over
     M_{a_0}-u-i M_{a_0}\Gamma_{a_0}   }+  excited,
\ea
where $B$ is given by the "box" diagram:
\be
B= 12 \left(\frac{m_u}{F_\pi}\right)^2Z^{-1}[R_{11}R_{12}+\ldots]
\ee
where dots stand for the contribution from diagrams with form factors, 
and $R_{ij}$ are taken from Table~\ref{isoscMixTab} (for $\eta$ and $\hat\eta$).
The coefficients $c_{\sigma\varphi\varphi}$ represent
the amplitudes describing decays of a scalar to a couple of
pseudoscalars; the calculation of them 
is discussed in the previous section. 
In general, they are momentum-dependent.

The kinematic invariants $s$, $t$ and $u$ are Mandelstam variables:
$s=(p_{\pi_1}+p_{\pi_2})^2$, $t=(p_{\eta}+p_{\pi_1})^2$,
$u=(p_{\eta}+p_{\pi_2})^2$

The ``{\it excited\/}''
terms are the contributions from excited scalar resonances
of a structure similar to that for the ground states.
The decay widths of  processes $\hat\eta\to \eta\pi\pi$ and
$\hat\eta'\to\eta\pi\pi$ are thereby
\be
\Gamma_{\hat\eta\to\eta\pi\pi}\approx 4\mev,\quad
\Gamma_{\hat\eta'\to\eta\pi\pi}\approx 6\mev.
\ee

For the processes $\hat\eta\to K\bar K\pi$
and $\hat\eta' \to K\bar K\pi$ we
approximate their decay widths by neglecting the pole-diagram
contribution
because it turns out that the "box" is dominant here.
The result is given in Table~\ref{etadecay}.

Unfortunately, the branching ratios for different decay modes 
of $\eta(1295)$ and $\eta(1440)$ are not known well from
experiment; so one can only find their total decay widths
\be
\Gamma^{\rm tot}_{\eta(1295)}=53\pm6 \mev,\quad
\Gamma^{\rm tot}_{\eta(1440)}=50-80 \mev,
\ee
which is in satisfactory agreement  with our results.

Strong and electromagnetic decays of the ground states of $\eta$
and $\eta'$ mesons 
were already investigated withing the framework of 
the standard NJL model in \cite{volk} and we do not consider them
here.

\section{Discussion and Conclusion}
Let us shortly recall some problems concerning the
interpretation of  experimental data on scalar and
$\eta$, $\eta'$ mesons. Several years ago,
attempts were undertaken to consider the state $\eta'(1440)$ as
a glueball \cite{geras}. There is an analoguous problem
with the interpretation of scalar states $f_0(1500)$ and $f_0(1710)$.
Moreover, the experimental
     status of the lightest scalar isoscalar singlet
     meson was unclear. In some papers, the resonance
     $f_0(1370)$ was considered as a member of the ground nonet
     \cite{F1370}, and it was not until 1998 that the resonance
     $f_0(400-1200)$ was included into the
     summary tables of PDG review%
     \footnote{ However, in earlier editions
     of PDG  the light $\sigma$ state still could be
     found;  it was excluded later.  }
     \cite{PDG}.

One will find a problem of the same sort in the case of $\kstar$.
The strange meson
     $K^*_0(1430)$ seems too heavy to be the ground state: $1\gev$
     is more characteristic of the ground meson states (see \cite{ishida,scadron}).

From our calculations we conclude that the states $\eta(1295)$
and $\eta(1440)$ can be considered as radial excitations of the
ground states $\eta$ and $\eta'$.
The calculation of their strong decay widths also confirms
 our conclusion. Let us note that these meson states are
significantly mixed.

In \cite{geras} the authors came to similar conclusions
about $\eta(1295)$ and $\eta(1440)$ where  the radial
excitations of the mesons were investigated in the
potential ${}^3P_0$ model.

Our calculations  also showed
that we can interpret the scalar states $f_0(1370)$, $a_0(1450)$, $f_0(1710)$
and $\kstar(1430)$
as the first radial excitations of $f_0(400-1200)$, $a_0(980)$, $f_0(980)$ and
$\kstar(960)$.
We estimated their masses and the widths of main decays in the framework of a nonlocal
chiral quark model.
We would like to emphasize that we did not use
  additional parameters except those necessary to fix the
mass spectrum of pseudoscalar mesons. We used the same form factors both
for the scalar and pseudoscalar mesons, which is required by the global chiral symmetry.

We assumed that the state $f_0(1500)$ is  a glueball, and its probable mixing
with $f_0(980)$, $f_0(1370)$ and $f_J(1710)$ may provide us with a
more correct description of the masses of these states%
\footnote{
Our estimates for the masses of $f_0$ and $\hat f_0$:
$M_{f_0}=1070\mev$ and $M_{\hat f_0}=1600\mev$ are expected to shift
to $M_{f_0}=980\mev$ and $M_{\hat f_0}=1710\mev$ after mixing with
the glueball $f_0(1500)$.
}%
(see Table~\ref{masses} and \cite{glueball}).  We are going to consider this problem in a
subsequent publication.

A more complicated situation takes place for the ground state $a_0(980)$.
In the framework of our quark-antiquark model, we have a mass deficit for this meson,
$830 \mev$ instead of $980\mev$.
We suspect that this drawback is due to a four-quark
component in this state which we did not take into account \cite{achasov}.

In future we are going to consider  glueball states \cite{glueball} and
to develop a model with quark confinement \cite{conf} to describe the momentum depentence
of meson amplitudes.

\section*{Acknowledgment}
We are very grateful to Prof. S.B.~Gerasimov for useful discussion.
This work has been supported by RFBR Grant N 98-02-16135.

\appendix

\section{Coefficients of the free part of effective Lagrangian
for the scalar isoscalar mesons.}

The functions $K_{\sigma(\varphi),ij}^{[a,a]}$ introduced in Sec.~4 (\ref{Lisosc})
are defined as follows
\ba
     K_{\sigma(\varphi),11}^{[a,a]}(P)&=&
     Z_{\sigma(\varphi),1}^a (P^2 -
     (m_q^a\pm m_{q'}^a)^2-  M_{\sigma^a(\varphi^a),1}^2 ),\nonumber\\
     K_{\sigma(\varphi),22}^{[a,a]}(P)
	 &=& Z_{\sigma(\varphi), 2}^a (P^2 -
     (m_q^a\pm m_{q'}^a)^2-  M_{\sigma^a(\varphi^a),2}^2 ),
     \nonumber \\
     K_{\sigma(\varphi),12}^{[a,a]}(P) &=& K_{\sigma(\varphi),21}^{[a,a]}(P) \;\; = \;\;
     \gamma_{\sigma(\varphi)}^a (P^2 - (m_q^a\pm m_{q'}^a)^2 ),\\
     K_{\sigma(\varphi),11}^{[8,9]}(P)&=&K_{\sigma(\varphi),11}^{[9,8]}(P)=
	\frac12 \left(T^{S(P)}\right)^{-1}_{89},\nonumber\\
	K_{\sigma(\varphi),12}^{[8,9]}(P)&=&
	K_{\sigma(\varphi),12}^{[9,8]}(P)=
	K_{\sigma(\varphi),21}^{[8,9]}(P)=0,\nonumber\\
	K_{\sigma(\varphi),21}^{[9,8]}(P)&=&
	K_{\sigma(\varphi),22}^{[8,9]}(P)=0,
	K_{\sigma(\varphi),22}^{[9,8]}(P)=0 \nonumber
\ea
where the ``bare'' meson masses are
\ba
     && M_{\sigma^8(\varphi^8),1}^2= (Z^8_{\sigma(\varphi),1})^{-1}
	\left({1\over 2}(T^{S(P)})^{-1}_{88} - 8I_1(m_u) \right),
     \nonumber \\
     && M_{\sigma^9(\varphi^9),1}^2= (Z^9_{\sigma(\varphi),1})^{-1}
	\left({1\over 2}(T^{S(P)})^{-1}_{99}-8I_1(m_s) \right), \nonumber\\
     && M_{\sigma^8(\varphi^8),2}^2=(Z^8_{\sigma(\varphi),2})^{-1}
	\left({1\over 2G} - 8I_1^{ff}(m_u) \right),\label{Mpuu}\\
     && M_{\sigma^9(\varphi^9),2}^2=(Z^9_{\sigma(\varphi),2})^{-1}
	\left({1\over 2G}-8I_1^{ff}(m_s) \right). \nonumber     
\ea
In the case of isoscalar mesons it is convenient to combine the scalar and pseudoscalar
fields into 4-vectors
\be
     \Sigma=(\sigma_{1}^{8\,r},\sigma_{2}^{8\,r},
	\sigma_{1}^{9\,r},\sigma_{2}^{9\,r}),\qquad
     \Phi=(\varphi_{1}^{8\,r},\varphi_{2}^{8\,r},
	\varphi_{1}^{9\,r},\varphi_{2}^{9\,r}), 
\ee
and introduce  $4\times 4$ matrix functions ${\cal K}_{\sigma(\varphi),ij}$,
instead of old $K_{\sigma(\varphi),ij}^{[a,b]}$,
where indices $i,j$ run from 1 through 4. This allows us to
rewrite the free part of the effective Lagrangian, which then, with
the meson fields renormalized, looks
as follows
\ba
     {\cal L}^{(2)}(\Sigma,\Phi)=\frac12\sum_{i,j=1}^4
     \left(\Sigma_i {\cal K}_{\sigma,ij}(P)\Sigma_j+
     \Phi_i {\cal K}_{\varphi,ij}(P)\Phi_j\right).
	\label{newL2}
\ea
and the functions  ${\cal K}_{\sigma(\varphi),ij}$ are
\ba
     {\cal K}_{\sigma(\varphi),11}(P)&=&P^2-(m_u\pm m_u)^2-M_{\sigma^8(\varphi^8),1}^2,\nonumber\\
     {\cal K}_{\sigma(\varphi),22}(P)&=&P^2-(m_u\pm m_u)^2-M_{\sigma^8(\varphi^8),2}^2,\nonumber\\
     {\cal K}_{\sigma(\varphi),33}(P)&=&P^2-(m_s\pm m_s)^2-M_{\sigma^9(\varphi^9),1}^2,\nonumber\\
     {\cal K}_{\sigma(\varphi),44}(P)&=&P^2-(m_s\pm m_s)^2-M_{\sigma^9(\varphi^9),2}^2,\\
     {\cal K}_{\sigma(\varphi),12}(P)&=&{\cal K}_{\sigma(\varphi),21}(P)=\Gamma_{\sigma_u(\eta_u)}(P^2-(m_u\pm m_u)^2),\nonumber\\
     {\cal K}_{\sigma(\varphi),34}(P)&=&{\cal K}_{\sigma(\varphi),43}(P)=\Gamma_{\sigma_s(\eta_s)}(P^2-(m_s\pm m_s)^2),\nonumber\\
     {\cal K}_{\sigma(\varphi),13}(P)&=&{\cal K}_{\sigma(\varphi),31}(P)=
	(Z^8_{\sigma(\varphi),1}Z^9_{\sigma(\varphi),2})^{-1/2}(T^{S(P)})^{-1}_{89}. \nonumber
\ea
Now,  to transform (\ref{newL2}) to conventional form, one
should just diagonalize a 4-dimensional matrix, which is better to
do numerically.

\section{The calculation of the amplitudes for the decays of the
excited scalar meson {$\hat a_0$}}
Here we collect some instructive formulae, which display a part of
details of the calculations made in this work. Let us demonstrate
how the amplitude of the decay $\hat a_0\to\eta\pi$ is obtained.
The mixing coefficients are taken from  Table~\ref{mixingTable}.
Moreover, the diagrams where  pion vertices contain form factors are
neglected because, as one can see from Table~\ref{mixingTable},
their contribution is significantly reduced.
\ba
      T^{(1)}_{\hat a_0\to\eta\pi}&=&4 \frac{m_u^2}{F_\pi}\biggl\{
	0.82\cdot0.71\cdot Z^{-1/2}\frac{I_2(m_u)}{I_2(m_u)}-\nonumber\\
	&&\left(1.17\cdot 0.71\cdot Z^{-1/2}-0.82\cdot 0.11\right)
	\frac{I_2^f(m_u)}{\sqrt{I_2(m_u)I_2^{ff}(m_u)}}-\nonumber\\
	&&1.17\cdot0.11\cdot\frac{I_2^{ff}(m_u)}{I_2^{ff}(m_u)}\biggr\}\approx0.2\gev,
\ea
\ba
     T^{(2)}_{\hat a_0\to\eta\pi}&=&
     2\frac{m_u^2}{F_\pi}(M_{a_0}^2-M_{\eta}^2-M_{\pi}^2)
     \biggl\{
        0.82\cdot0.71 Z^{-1/2}\frac{I_3(m_u)}{I_2(m_u)}-\nonumber\\
	&&\left(1.17\cdot 0.71\cdot Z^{-1/2}-0.82\cdot 0.11\right)
	\frac{I_3^f(m_u)}{\sqrt{I_2(m_u)I_2^{ff}(m)}}-\nonumber\\
	&&1.17\cdot0.11\frac{I_3^{ff}(m_u)}{I_2(m_u)}\biggr\}\approx3.5 \gev.
\ea
The decay width  thereby is
\be
     \Gamma_{\hat a_0\to\eta\pi}=
     \frac{|T_{\hat a_0\to\eta\pi}|^2}{16\pi M_{\hat a_0}^3}
     \sqrt{M_{\hat a_0}^4\!+\!M_{\eta}^4\!+\!M_{\pi}^4\!-\!
     2(M_{\hat a_0}^2 M_{\eta}^2\!+\!M_{\hat a_0}^2 M_{\pi}^2\!+\!M_{\eta}^2 M_{\pi}^2)}\approx
     160 \mev.
\ee
Here $I_2(m_u)=0.04$, $I_2^f(m_u)=0.014c$, $I_2^{ff}(m_u)=0.015c^2$,
$I_3(m_u)=0.11 \gev^{-2}$, $I_3^f(m_u)=0.07c\gev^{-2}$,$I_3^{ff}(m_u)=0.06c^2\gev^{-2}$
and $c$ is the external form factor parameter factored out
and cancelled in the ratios of the integrals.

For the decay into strange mesons we obtain (see Fig.1)
\ba
     &&\!\!\!\!T_{\hat a_0\to K^+K^-}\!=\!
     C_K\!\left(-\frac{iN_c}{16\pi^2}\right)\!\!\int\!d^4k
	{ \Tr[(m_u+\Slash{k}+\Slash{p}_1)\gamma_5(m_s+\Slash{k})\gamma_5(m_u+\Slash{k}-\Slash{p}_2)]
      \over
	(m_s^2-k^2)(m_u^2-(\Slash{k}-\Slash{p}_1)^2)(m_u^2-(\Slash{k}-\Slash{p}_2)^2)
	}\approx\nonumber\\
     &&\!\!\!\!2C_K\left\{
	 (m_s+m_u)I_2(m_u)-\Delta I_2(m_u,m_s)-
	 [m_s(M_{\hat a_0}^2-2M_{K}^2)-\right.\\
     &&\!\!\!\!\left.\quad	2\Delta^3]I_3(m_u,m_s)\nonumber
        \right\},
\ea
where $\Delta=m_s-m_u$ and
\be
	I_3(m_u,m_s)=-i\frac{N_c}{(2\pi)^4}\int_{\Lambda_3}\!\frac{d^4k}{(m_u^2-k^2)^2(m_s^2-k^2)}.
\ee
The coefficient $C_K$ absorbs the Yukawa coupling constants and some structure coefficients.
The integral $I_2(m_u,m_s)$ is defined by (\ref{DefIf}).
This is only the part of the amplitude without form factors.
The complete amplitude of this process is a sum of contributions which
contain also the integrals $I_2^{f..f}$ and $I_3^{f..f}$ with form factors.
Thus, the amplitude is
\ba
	&&T_{\hat a_0\to K^+K^-}=T^{(1)}+T^{(2)},\\
	&&T^{(1)}=\frac{m_u+m_s}{2F_K}
	\bigl\{
	(m_s+m_u)\cdot 0.13-\Delta\cdot 0.21\bigr\}\approx 0.2\gev,\\
	&&
	T^{(2)}=\frac{m_u+m_s}{2F_K}\bigl\{[m_s(M_{a_0}^2-2M_{K}^2)-2\Delta^3]\cdot1\gev^{-2}\bigr\}\approx 2.3\gev
	,\\
	&& F_K=1.2F_\pi.\nonumber
\ea
The decay width therefore is evaluated to be
\be
	\Gamma_{\hat a_0\to K^+K^-}=\Gamma_{\hat a_0\to \bar K^0K^0}\approx 50 \mev.
\ee


\begin{table}[p]
\caption{The mixing coefficients for the ground and first radially excited states of
the scalar and pseudoscalar isovector and strange mesons.
The caret symbol marks the excited states.}
\label{mixingTable}
$$
	\begin{array}{||r|c|c||}
	\hline
		& a_0 & \hat a_0\\
	\hline
	a_{0,1}	&0.87	  &0.82    \\
	a_{0,2} &0.22	  &-1.17   \\
	\hline
	\end{array}
\quad
	\begin{array}{||r|c|c||}
	\hline
		& \kstar & \hat\kstar\\
	\hline
	\kstar{}_{,1}	&0.83	  &0.89    \\
	\kstar{}_{,2} &0.28	  &-1.11   \\
	\hline
	\end{array}
$$
$$
	\begin{array}{||r|c|c||}
	\hline
		& \pi & \hat\pi\\
	\hline
	\pi_1	&1.00	  &0.54    \\
	\pi_2   &0.01	  &-1.14   \\
	\hline
	\end{array}
\quad
	\begin{array}{||r|c|c||}
	\hline
		& K & \hat K   \\
	\hline
	K_1	&0.96	  &0.56   \\
	K_2    & 0.09	  &-1.11  \\
	\hline
	\end{array}
$$

\end{table}

\newpage
\begin{table}[p]
\caption{The mixing coefficients for the isoscalar meson states}
\label{isoscMixTab}
$$
	\begin{array}{||r|c|c|c|c||}
	\hline\hline
	 		&\eta 		&\hat\eta 	&\eta' 		&\hat\eta'\\
	\hline
	\varphi^8_{1}	&0.71		&0.62		&-0.32		&0.56		\\
	\varphi^8_{2}   &0.11		&-0.87		&-0.48		&-0.54		\\
	\varphi^9_{1}	&0.62		&0.19		&0.56		&-0.67		\\
	\varphi^9_{2}   &0.06		&-0.66		&0.30		&0.82		\\
	\hline
	\end{array}
$$
$$
	\begin{array}{||r|c|c|c|c||}
	\hline\hline
	 		&\sigma 	&\hat\sigma 	&f_0 		&\hat f_0\\
	\hline
	\sigma^8_{1}	&-0.98		&-0.66		&0.10		&0.17		\\
	\sigma^8_{2}&0.02		&1.15		&0.26		&-0.17		\\
	\sigma^9_{1}	&0.27		&-0.09		&0.82		&0.71		\\
	\sigma^9_{2}&-0.03		&-0.21		&0.22		&-1.08		\\
	\hline
	\end{array}
$$
\end{table}
%
\newpage

\begin{table}[p]
\caption{The model masses of mesons, MeV}
\label{masses}
$$
\begin{array}{||l|c|c|c|c||}
	\hline \hline
		& GR 		& EXC		& GR(Exp.)\,\cite{PDG}		&EXC(Exp.)\,
\cite{PDG} 	\\
	\hline
M_{\sigma} 	& 530		& 1330		&400-1200		&1200-1500	\\
M_{f_0}		& 1070		& 1600		&980\pm10		&1712\pm5	\\
M_{a_0}		& 830		& 1500		&983.4\pm0.9		&1474\pm19	\\
M_{\kstar} 	& 960		& 1500		&905\pm50\,\cite{ishida}&1429\pm12	\\
M_{\pi} 	& 140		& 1300		&139.56995\pm0.00035	&1300\pm100	\\
M_{K} 		& 490		& 1300		&497.672\pm0.031	&1460(?)	\\
M_{\eta} 	& 520		& 1280		&547.30\pm0.12		&1297.8\pm2.8	\\
M_{\eta'} 	& 910		& 1470		&957.78\pm0.14		&1440-1470	\\
	\hline \hline
\end{array}
$$
\end{table}

\newpage
\begin{table}[p]
\caption{$\eta(1295)$ and $\eta(1440)$ decay modes. }
\label{etadecay}
\begin{tabular}{||c|c|c|c|c|c|c||}
\hline\hline
 &   $a_0\pi$ & $\eta\sigma$ & $\eta\pi\pi$ & $K\bar K\pi$ & $KK^*$&
$\Gamma_{\rm tot}$\\
\hline
$\eta(1295)$ &  $3\mev$  & $30\mev$  & $4\mev$& $5\mev$ & $-$ & 48\mev \\
\hline
$\eta(1440)$ &  $10\mev$  & $3\mev$  & $6\mev$& $26\mev$ & $70~{\rm keV}$ &45\mev\\
\hline
\end{tabular}
\end{table}

\newpage
\centerline{ FIGURE CAPTIONS}
\begin{itemize}
\item[1)]  Diagrams describing decays of $\hat a_0$ to pseudoscalars.
\item[2)]  Diagrams describing the decay  $\hat\eta\to\eta\pi\pi$. The black box stands for
the sum of ``box'' diagrams represented by one-loop quark graphs with four meson vertices.
The rest of the diagrams is a set of pole graphs with the $\sigma$, $f_0$ and $a_0$ scalar resonances.
 The diagram with $a_0$ is to be taken into account for two channels (due to exchange of the pions momenta).
 There are analogous contributions from radially excited resonances.
\end{itemize}

\newpage
\begin{figure}[h]
\begin{center}
\psfig{file=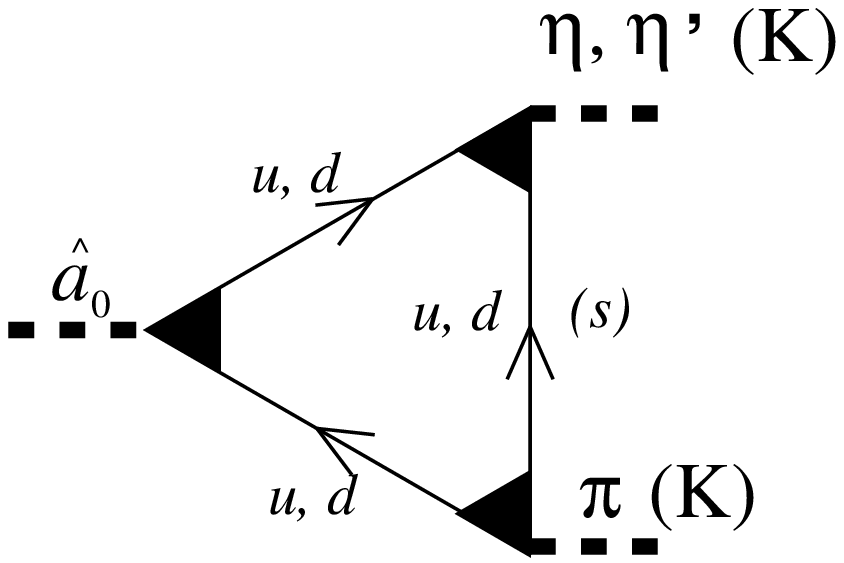}
\caption{ }
\end{center}
\end{figure}

\newpage
\begin{figure}[h]
\begin{center}
\psfig{file=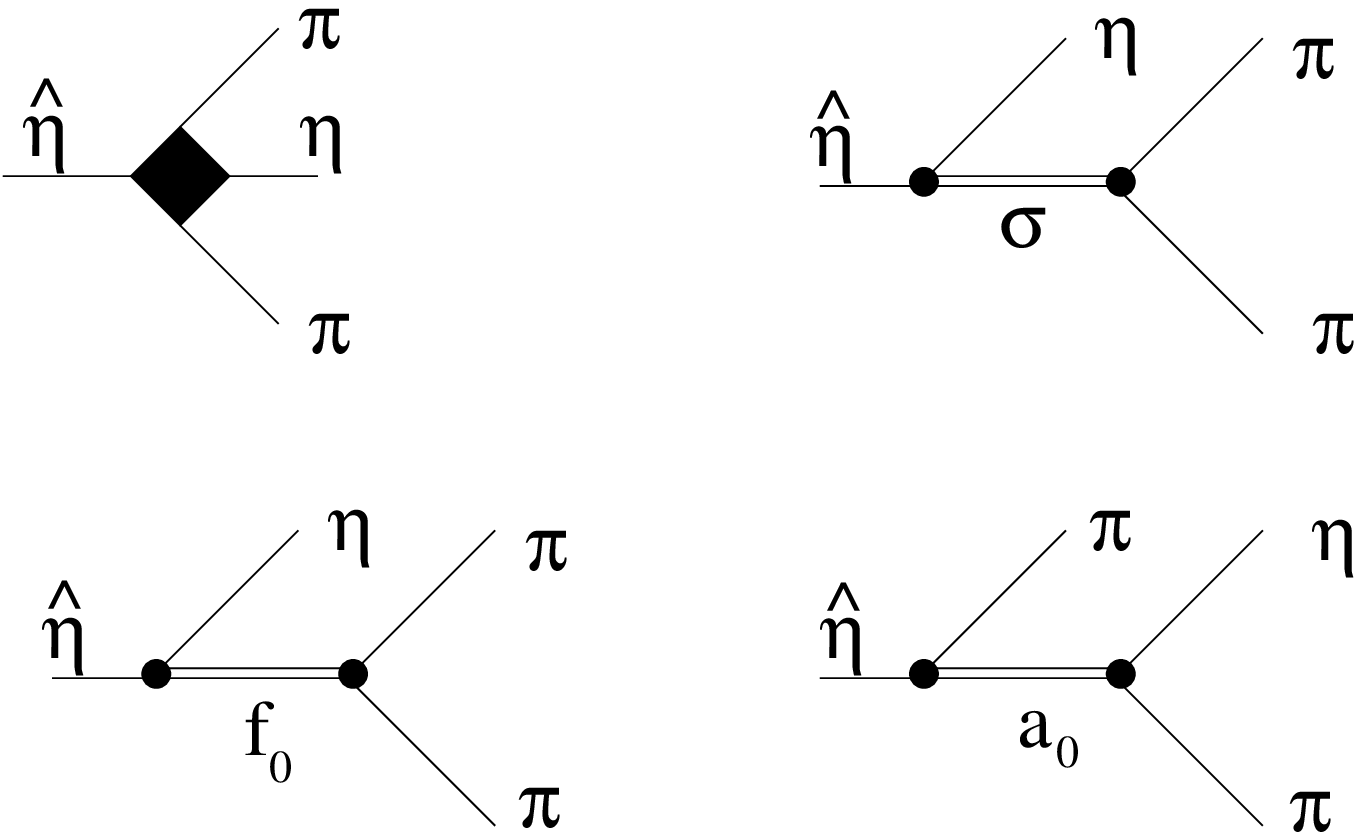}
\caption{   }
\end{center}
\end{figure}

\end{document}